\documentclass[fleqn,usenatbib]{mnras}

\usepackage{newtxtext,newtxmath}

\usepackage[T1]{fontenc}
\usepackage{ae,aecompl}
\usepackage{graphicx}
\usepackage{booktabs}
\usepackage{multirow}
\usepackage{xspace}
\usepackage{comment}
\usepackage{siunitx}
\usepackage{overpic}
\DeclareSIUnit{\eV}{eV}
\DeclareSIUnit{\keV}{keV}
\DeclareSIUnit{\parsec}{pc}
\DeclareSIUnit{\erg}{erg}
\DeclareSIUnit{\gauss}{G}
\DeclareSIUnit{\year}{yr}
\usepackage{hyperref}
\def\integral {\emph{INTEGRAL}\ }

\usepackage{natbib} 

\title[blz]{
INTEGRAL search for magnetar giant flares from the Virgo Cluster and in nearby galaxies with high star formation rate
}

\author[ ]{Dominik P. Pacholski$^{1,2}\thanks{E-mail: dominik.pacholski@inaf.it}$, Edoardo Arrigoni$^{1,2}$, Sandro Mereghetti$^{1}$,  Ruben Salvaterra$^{1}$ 
\\
$^{1}$ INAF, Istituto di Astrofisica Spaziale e Fisica Cosmica Milano, via A.\ Corti 12, I-20133 Milano, Italy\\
$^{2}$ Universit\`a degli Studi di Milano Bicocca, Dipartimento di Fisica G. Occhialini, Piazza della Scienza 3, 20126 Milano, Italy
}

\date{Accepted 2024 November 5. Received 2024 November 5; in original form 2024 September 18}

\pubyear{2024}

\begin{document}
\label{firstpage}
\pagerange{\pageref{firstpage}--\pageref{lastpage}}
\maketitle

%
%
 
  \begin{abstract}
Giant flares from magnetars can reach, for a fraction of a second, luminosities greater than  10$^{47}$ erg s$^{-1}$ in the hard X-ray/soft $\gamma$-ray range. This makes them visible at distances of several megaparsecs. However, at extragalactic distances (farther than the Magellanic Clouds) they are difficult to distinguish from the short $\gamma$-ray bursts, which occur much more frequently. Since magnetars are  young neutron stars, nearby galaxies with a high rate of star formation are optimal targets to search  for magnetar giant flares (MGFs). Here we report the results of a search for MGFs in  observations of the Virgo cluster   and in a small sample of nearby galaxies obtained with the  IBIS instrument on the INTEGRAL satellite. From the currently known MGF sample we find that their energy distribution is well described by a power law with slope $\gamma$=2 (with 90\% c.l. interval [1.7-2.2]). From the  lack of detections in this extensive data set (besides 231115A in M82)   we derive a 90\% c.l. upper limit on the rate of MGF with $E>3\times10^{45}$ erg of $\sim2\times10^{-3}$ yr$^{-1}$  per magnetar
and a lower limit of $R(E)>\sim4\times10^{-4}$ yr$^{-1}$ magnetar$^{-1}$ for $E<10^{45}$ erg.
\end{abstract}

   \maketitle
%

\section{Introduction} 

Magnetars are  young isolated neutron stars whose electromagnetic emission (mostly at energies between $\sim$0.1 and few hundreds keV) is powered by the dissipation of their intense 
magnetic fields (generally $>10^{14}$ G in the magnetosphere and up to $\sim10^{16}$ G in the NS interior). 
They exhibit  strong variability on all  timescales, from milliseconds to months  \cite[see, e.g.,][for reviews]{2015SSRv..191..315M,2017ARA&A..55..261K}. 
The most spectacular, although quite rare, variability events observed in these objects are the so called giant flares, during which a hard X-ray luminosity up to $\sim10^{47}$ erg s$^{-1}$ can be reached for a fraction of a second. The extreme properties  of the  first discovered giant flare, the famous event of March 5, 1979 \citep{1979Natur.282..587M,1980ApJ...237L...1C}, were among the main motivations to invoke  neutron stars endowed with ultra-strong magnetic fields, which are at the basis of the magnetar model \citep{1992ApJ...392L...9D,1992AcA....42..145P}. 

After the March 1979 event, that was produced by a magnetar in the Large Magellanic Cloud, only two other giant flares were observed, from two different Galactic magnetars (see Table~\ref{tab:MGF}). 
These three  magnetar giant flares  (MGFs) shared a common characteristic in their light curves: an initial short ($\lesssim$0.2-0.3 s), very energetic ($E \gtrsim 10^{46}$ erg) hard X-ray pulse, always followed by a long, fainter tail of softer emission,  periodically modulated at the magnetar spin period of a few seconds \citep{2008A&ARv..15..225M}.  These periodic tails are the telltale "signature" that allows us to recognise a MGF and to associate it with a known source, even in the lack of positional information, as it occurred for the MGF of 27 December 2004 from SGR 1806--20 \citep{2004GCN..2920....1B}.

It was soon realised that the luminous initial pulses of MGFs could be detected up to distances of few tens of Mpc, while the pulsed tails would remain undetectable. Thus a MGF from a source in a distant galaxy   appears very similar, and difficult to distinguish from, a short GRB \citep{1982Ap&SS..84..173M}. Several studies were carried out to determine the fraction of the short GRB population that could be due to extragalactic MGFs. However, the paucity of candidates makes the  resulting estimates quite uncertain, with reported values in the wide range from 1\% up to 40\%   \cite[see, e.g.,][]{2005Natur.434.1098H,2006ApJ...640..849N,2007ApJ...659..339O,2015MNRAS.447.1028S}.

\begin{table*}
    \centering
    \caption{Properties of the three confirmed MGFs  and of the candidates found in external galaxies.}
    \label{tab:MGF}
    \begin{tabular}{lccccc}
         \toprule
             \midrule
         Source & Galaxy  &  E$_\mathrm{GF}$ [$10^{45}$ erg] &  $\alpha$ & T [s] & References\\
                & d [Mpc]  & L$_\mathrm{peak}$ [$10^{46}$ erg s$^{-1}$] & $E_p$ [keV] &  & \\
\midrule          
\midrule        
790305         & LMC & 0.7 & -- & <0.25  & [3],[4]\\
SGR 0526$-$66  & 0.055  & 0.65 & 500 & &\\
\midrule
980827         & Milky Way & 0.43 & -- & <1.0 &[3],[4]\\
SGR 1900$+$14    & 0.0125  & 2.3& 1200 & & \\
\midrule
041227         & Milky Way &  7.7 & -0.7 & 0.18 &[1], [2]\\
SGR 1806$-$20    & 0.0087  &  12 & 850 & &\\
\midrule\midrule
051103         & M81  & 53 & -0.3 & 0.14 & [6]\\
         &       3.7 & 180 &  2300& &\\
\midrule
070201         & M31  & 1.5 & -0.98 & 0.256 & [3]\\
     &     0.78 & 12 & 296 & &\\
\midrule
070222          & M83 & 6.2 & -1.0 & 0.038 & [4]\\
     &  4.5  &  40 & 1290 & &\\
\midrule
180128A          &  NGC 253 & 0.60 & 0.6 & 0.2 & [8]\\
     &       3.5 & 11 & 290 & &\\
\midrule
200415A          & NGC 253  & 13 & 0.0 & 0.10 & [5],[6]\\
     &     3.5 & 140 & 887 & &\\
\midrule
231115A          & M82 & 1 & 0.04 & 0.093 &[7],[9]\\
     &      3.6 & 5.1 & 551 & &\\
\bottomrule\\[-5pt]
\end{tabular}

\raggedright
Isotropic energy $E_{GF}$ and peak luminosity L$_\mathrm{peak}$ are in the range of 1 keV - 10 MeV.
References: [1] \citet{2007AstL...33....1F}, [2] \citet{2008MNRAS.386L..23B}, [3] \citet{2008ApJ...680..545M}, [4] \citet{2021ApJ...907L..28B}, [5] \citet{2021Natur.589..207R}, [6] \citet{2021Natur.589..211S}, [7] \citet{2024Natur.629...58M}, [8] \citet{2024A&A...687A.173T}, [9] \citet{2024arXiv240906056T}.

\end{table*}

\begin{table}
    \centering
    \caption{Nearby galaxies with high star formation rate}
    \label{tab:galaxies}
    \begin{tabular}{lccc}
         \toprule
             \midrule
 Galaxy  &  Distance & SFR &   Exposure \\
        & [Mpc]  & [$M_\odot$ yr$^{-1}$] & [Ms]  \\
\midrule          
\midrule        
NGC 253         & 3.5  & 4.9  & 0.62   \\
M81          & 3.7   & 0.5  &  25.5  \\
M82        &   3.6 & 7.1  &  26.1  \\
M83         &  4.5  & 4.2  &  5.2  \\
NGC 4945        &  3.4  & 1.45  &  17.9  \\
IC 342           &  2.3  & 1.9  &  7.2  \\
PGC 50779           & 4.2   & 3.9  &  20.8 \\

\bottomrule\\[-5pt]     
\end{tabular}
\end{table}

 \begin{table}
    \centering
        \caption{Coefficients of proportionality between rate of core collapse supernovae (in units of SN/century) and galaxy luminosity (in units of $10^{10} L_\odot$) assumed in this work. See \citet{2015MNRAS.447.1028S}.}
    \begin{tabular}{llc}
             \hline
             \hline
        Galaxy type & & $k_\mathrm{morph}$ \\
        \hline
        \hline
        Elliptical & & 0.05 \\
        Spiral  & S0 & 0.05\\
         Spiral       & Sa-b & 0.89\\
         Spiral       & Sc-d & 1.68\\
         Spiral       & Sm & 1.46\\
         Spiral       & unclassified & 1\\
        Irregular & & 1.46\\
        Dwarf & & 0\\
                \hline

    \end{tabular}
    \label{tab:kmorphvalues}
\end{table}

A few candidate MGFs have been identified among short GRBs with positions consistent with bright galaxies outside the Local Group.  
Although an MGF origin cannot be proven with absolute certainty in all these cases, at least some of them appear very plausible \citep{2021Natur.589..211S,2021Natur.589..207R,2024Natur.629...58M}. Furthermore, significant evidence for a population of extragalactic MGFs emerges from a collective statistical analysis of these candidates \citep{2021ApJ...907L..28B}.
Their main properties are compared to  those of the three confirmed MGFs in Table~\ref{tab:MGF}.

During giant flares, magnetars emit a significant fraction of their total magnetic energy budget. This limits the number of the most energetic events that can be emitted in a magnetar's lifetime.
It is therefore interesting to derive observational constraints on the rate of MGFs as a function of energy.
Given the low rate of  Galactic MGFs, it is clear that targeted searches in more distant galaxies could be a promising way to enlarge the sample  and    derive constraints on the MGFs rate of occurrence. Regions of high star formation rate (SFR) offer the best prospects, since magnetars are believed to originate from massive stars undergoing core-collapse supernova explosions. 
Here we report on a search for MGFs in INTEGRAL observations of the Virgo cluster and in a few other nearby galaxies with high SFR.

\begin{figure*}
    \centering
    \includegraphics[width=1\linewidth]{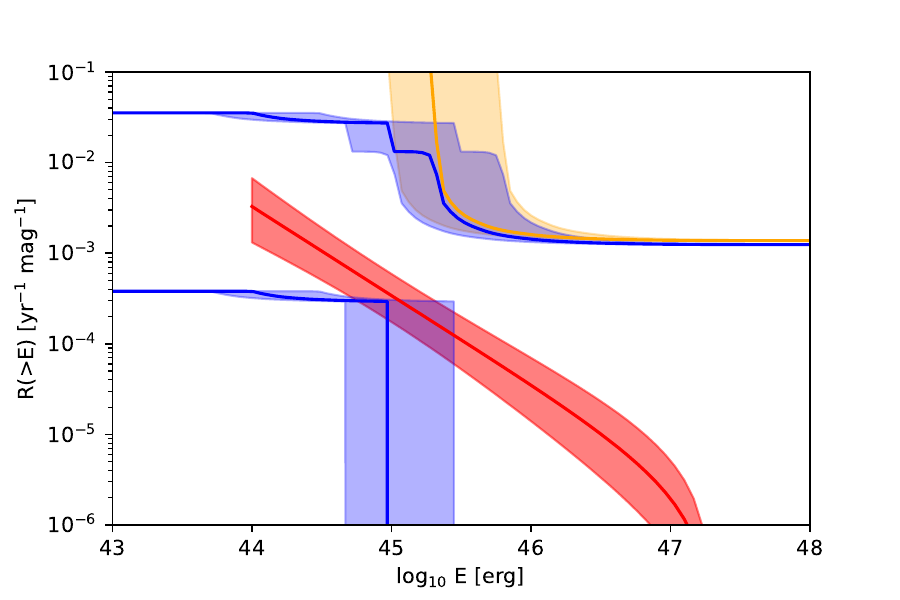} 
    \caption{Rate of magnetar giant flares as a function of energy. The red line is the distribution we derived with a maximum likelihood analysis of  seven MGF with fluence above $2\times 10^{-6}$ erg cm$^{-2}$ observed in the last 30 yrs. The shaded area indicates the 90\% c.l. interval. 
    The upper limit (90\% c.l.) derived from the INTEGRAL observations of the Virgo cluster of galaxies is indicated by the orange line. 
    The blue lines give the lower and upper limits with the inclusion of seven nearby galaxies with high star formation rate.
    The shaded blue and orange areas indicate the uncertainty resulting from the   spectral parameters assumed for the MGF.    
    }
    \label{fig:GFendistr}
\end{figure*}

\section{Data analysis and results}
 
We use data obtained with the IBIS instrument on board INTEGRAL.
IBIS is a coded-mask imager composed of two detectors, ISGRI and PICsiT, covering the 15 keV — 10 MeV energy range. 
In this work, we   use only the data of ISGRI  \citep{2003A&A...411L.141L}, which operates in the nominal $15-1000$ keV  range,  providing timing information  with a  resolution of   61 $\mu$s for each detected count. Thanks to its  wide field of view,  $29^{\circ}\times29^{\circ}$ (with a central $\sim9^{\circ}\times9^{\circ}$ region at full sensitivity) it allows to simultaneously monitor several sources and is well suited to observe the Virgo cluster which extends over a sky region centred at R.A.=188$^{\circ}$, Dec.=12$^{\circ}$ with  a radius of $\sim8^{\circ}$. The  angular resolution of ISGRI is $12'$,  but sources can be located with an accuracy of a few arcminutes.

To optimise the imaging performance, INTEGRAL points to the sky according to a predefined dithering pattern. Therefore, the  data are split into many  short pointings with fixed attitude and typical duration between 30 and 60 minutes each, called science windows (ScWs), interleaved by short slews.  As a first step, we searched for bursts by analysing the light curves of all the individual ScWs selected as described below.
All the burst candidates found in this search were then checked with an interactive imaging analysis, in order to  eliminate events caused by background variations, instrumental effects (e.g. noisy pixels) and sources unrelated to the galaxies under study. 

For the light curves analysis we followed the procedure described in \citet{2021ApJ...921L...3M}. Briefly, this consists in evaluating the background level of each ScW through an iterative 3$\sigma$ clipping, and then searching for count rate excesses on different time scales. We used a threshold corresponding to a false alarm probability of 10$^{-3}$ in each ScW. 
The ScWs with strongly variable background over their whole duration were excluded from the burst search. This resulted in a reduction of less than 4\% of the total exposure time.
The imaging analysis was carried out with the software developed for the Integral Burst Alert System (IBAS, \citealp{2003A&A...411L.291M}).

\subsection{Virgo cluster}

At a distance of  $\sim$16.5 Mpc, Virgo is the closest cluster of galaxies and it   has been extensively observed by INTEGRAL. 
We considered all the galaxies reported in the Extended Virgo Cluster Catalog (EVCC, \citealp{2014ApJS..215...22K}),  excluding the dwarf galaxies. 
We then extracted from the \integral archive all the ScWs covering the positions of the 975 selected galaxies.
This resulted in  10,886 ScWs between May 2003 and August 2023, yielding  a total  observation time of 34.8 Ms.  On average, a single galaxy was observed for 22 Ms in 6700 ScWs. 
Since in general each ScW contains several galaxies, we did not apply any spatial selection and extracted light curves using the counts from the whole detection plane of ISGRI.
This was done in the nominal 15-300 keV energy range in eight integration times with logarithmically spaced durations between 0.01 and 1.28 s.  
Statistically significant count rate peaks found in adjacent time bins and/or in overlapping timescales were grouped and considered as a single  burst candidate.
All the burst candidates  were then examined by producing the sky images of the corresponding time intervals, but none of them  could be confidently interpreted as a real burst from one of the Virgo galaxies in the field of view. Note that at the Virgo cluster distance, the pulsed tail of a MGF would produce a hard X-ray fluence  of the order of 10$^{-9}$-10$^{-8}$ erg cm$^{-2}$ over a time interval of few hundreds seconds. Since this is below the sensitivity of ISGRI, we concentrated on the search for short duration events.

\subsection{Nearby galaxies}

We considered the seven nearby galaxies with high SFR listed in Table \ref{tab:galaxies}. 
To search for bursts, we applied the same procedure described above, with the only difference that the   light curve of each galaxy was extracted selecting only the ISGRI detector pixels illuminated by the source for more than 50\% of their surface. 
Also in this case no significant bursts from the considered galaxies were found.

\section{Discussion}

The results of our search for  MGFs in the INTEGRAL data can be used to constrain the rate of occurrence of these events.
To do this, we need the energy distribution of MGFs and an estimate of the expected numbers of MGFs in our data sets. 

\subsection{Energy distribution of MGFs}

Given that the energy   distribution of MGFs is only poorly constrained by the three   confirmed events seen up to now, we assume that it can be described by a power law  with index $\gamma$ in the range [E$_{min}$,E$_{max}$] and estimate the most likely parameters using a larger set of events selected among those listed  in Table~\ref{tab:MGF} as described below.

The expected number of MGF with energy in the range [E$_1$,E$_2$] detectable in an observation of duration $T$ is given by

\begin{equation}
    \mu_{12} = \tilde{k}\frac{\gamma -1}{E_{min}^{1-\gamma}-E_{max}^{1-\gamma}}\int_{E_{1}}^{E_{2}}E^{-\gamma} N(E) T \mathrm dE,
     \label{eq:mu}
\end{equation}

\noindent
where the normalisation $\tilde{k}$ represents the total number of MGF per year per magnetar and $N(E)$ is 
the number of magnetars from which an event of energy $E$ would be detectable.

As discussed in  \citet{2021ApJ...907L..28B}, the whole sky was covered down to a limiting fluence of $S_\mathrm{IPN}=2\times 10^{-6}$ erg cm$^{-2}$ with the instruments of the InterPlanetarty Network (IPN) during the last 30 yrs.
Therefore, we set $T$=30 yrs in eq.~(\ref{eq:mu}) and select from Table~\ref{tab:MGF} the   MGFs with fluence above $S_\mathrm{IPN}$ observed in this time period. Only 790305 and 231115A do not satisfy these two requirements and thus  are excluded, leading to a complete sample of seven MGFs.

To estimate $N(E)$ we consider all the galaxies within $d_\mathrm{max} = \sqrt{E/4\pi S_\mathrm{IPN}}$  reported in the \textit{z0MGS} catalog \citep{2019ApJS..244...24L}, 
and assume that each of them contains a number of magnetars $N_i$ proportional to its SFR:  
$N_i = N_\mathrm{MW+MC} ~(SFR_i /  SFR_{MW+MC})$
= 30 ($SFR_i / ~1.85M_\odot$ yr$^{-1}$).
We have assumed 30 magnetars in the Milky Way and Magellanic Clouds, based on the number of  known magnetars. This is probably an underestimate due to the presence of still undiscovered quiescent magnetars. All the derived rates scale linearly with    $N_\mathrm{MW+MC}$.

We then apply a maximum likelihood method to determine the $\tilde{k}$ and $\gamma$ values that best reproduce the observed data,
considering that the observed number of MGFs with energy $E_1<E<E_2$ follows a Poisson distribution with average $\mu_{12}$.
In this way, we find $\gamma = 1.97$ and $\tilde{k} = 3.5\ 10^{-3}$ yr$^{-1}$   magnetar$^{-1}$, with  90\% c.l. intervals of [1.73,2.21] and [1.3,6.7] 10$^{-3}$, respectively.
The corresponding distribution, in its integral form, is shown by the red line in Fig.~\ref{fig:GFendistr}, where the shaded area indicates the 90\% c.l. uncertainty.

\subsection{Expected number of MGFs}

We can   calculate the expected number of MGFs in the INTEGRAL observations of the Virgo cluster by summing the contributions of all the individual galaxies: 

\begin{equation}
 N_\mathrm{GF} = \sum_i N^\mathrm{GF}_i = \sum_i \int T_i(<S) R_\mathrm{GF}(S)\ N_i  dS 
 \label{eq:Ngf}
\end{equation}

\noindent
where $R_\mathrm{GF}(S)$ is the rate of GF with fluence $S =  \frac{E}{4\pi d_i^2}$ emitted by a single magnetar.
We took $d_i$ = 16.5 Mpc for all the Virgo cluster galaxies. 
$T_i(<S)$ is the  total  time in which galaxy  $i$ has been observed with sensitivity better than $S$.
In the computation of  $T_i(<S)$, we  took into account  the dependence of the ISGRI sensitivity on the position in the field of view and on the background level measured in each ScW. 
The instrument sensitivity in the energy range used for the search  depends also on the spectral shape of the burst. 
We assume an exponentially cut-off power law spectrum ($F(E) = k E^{\alpha} {\rm exp}(-(E(2+\alpha)/E_p))$  ph cm$^{-2}$ s$^{-1}$ keV$^{-1}$).

We expect that the number of magnetars per galaxy is proportional to the rate $R_\mathrm{SN}$ of core collapse supernovae.
Therefore, we estimate it by scaling the number of Galactic magnetars:  $N_i = N_{MW} R_i^\mathrm{SN} / R_{MW}^\mathrm{SN}$. 
The rate of core collapse supernovae   depends on the luminosity and morphological type of each galaxy:
$R^{SN} = k_\mathrm{morph} L_g$.
We used the  $g$ band  luminosities $L_g$  reported in the EVCC and  the values of $k_\mathrm{morph}$   given in Table~\ref{tab:kmorphvalues}. 

As a representative spectral shape for a MGF initial spike, we take the parameters measured with INTEGRAL for 231115A in M82:  
$\alpha$=0.04 and $E_p$=551 keV \citep{2024Natur.629...58M}.
For these values,  eq.~\ref{eq:Ngf} gives an expected number of $N_\mathrm{GF} = 0.15$ in our  Virgo cluster observations. By varying the spectral parameters in the range observed in other MGFs ($\alpha$=[--1.0,0.6] and $E_p$=[300, 2300] keV), $N_\mathrm{GF}$ varies from 0.05 to 0.3.

\subsection{Constraints on the MGF rate}

Given that no MGF was detected in Virgo, we computed the rate upper limit as
\begin{equation}
    R_\mathrm{up}(S) = \frac{N_\mathrm{up}}{\sum_i T_i(<S)N_i},
\end{equation}
with $N_\mathrm{up}=2.303$, which corresponds to the 90\% c.l. \citep{1986ApJ...303..336G}. The corresponding limit on the rate  as a function of MGF energy $E$  is indicated by the orange line in Figure~\ref{fig:GFendistr}, where the shaded region reflects the uncertainty given by varying the spectral parameters in the range mentioned above.

Owing to the relatively large distance of the Virgo cluster, these observations cannot constrain the rate of MGFs with energy below a few 10$^{45}$ erg. 
On the other hand, this can be done with the results obtained for the seven galaxies of Table~\ref{tab:galaxies}, which globally are expected to contain less magnetars, but are closer than the Virgo cluster.  In this case, one MGF was detected with INTEGRAL (231115A in M82, \citealt{2024Natur.629...58M}). Therefore, using the formalism described above, we can put both an upper and a lower limit to the MGF rate. As indicated by the two blue lines in  Figure~\ref{fig:GFendistr},  the limits (at 90\% c.l.) obtained from these seven galaxies (plus those in the Virgo cluster) extend to lower values of $E$.

\begin{figure*}
    \centering
    \includegraphics[width=1\linewidth]{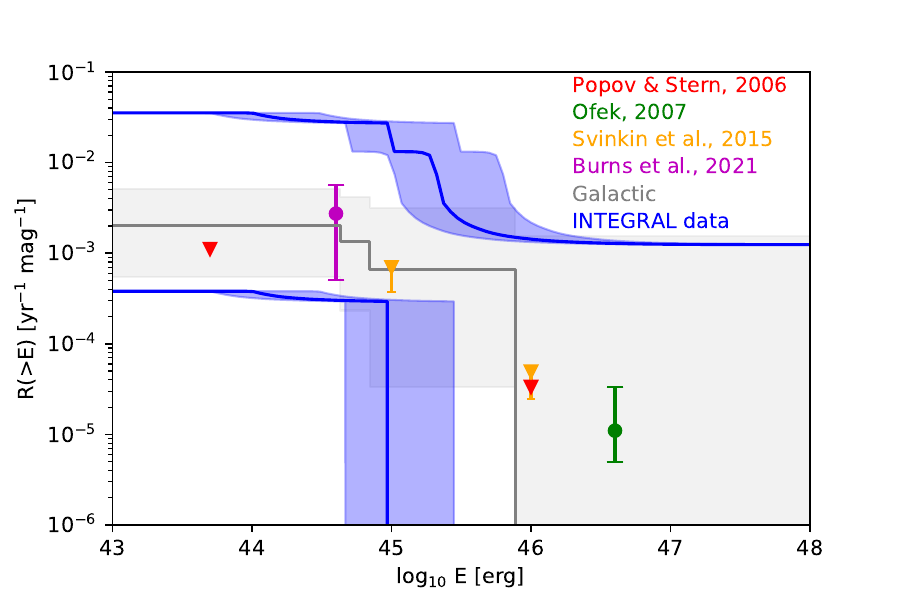} 
    \caption{Lower and upper limits (90\% c.l.) on the rate of MGFs as a function of emitted energy $E$ derived in this work (blue lines, with shaded uncertainty regions resulting from the uncertainties on MGF spectra). The coloured symbols give the    rate values (dots) or upper limits (triangles)  reported by different authors
    (rescaled to the same assumptions to allow a proper comparison, see text for details).
    The black histogram (with 90\% c.l.  uncertainty shaded) is the rate derived using only  the three confirmed MGFs (790305 in the LMC, 980827 and 041227 in our Galaxy).
        }
    \label{fig:GFbiblim}
\end{figure*}

In Fig.~\ref{fig:GFbiblim} we compare the limits obtained in our work with the  estimates derived by different authors  \citep{2006MNRAS.365..885P, 2007ApJ...659..339O,2015MNRAS.447.1028S,2021ApJ...907L..28B}. 
We converted all these measurements to the same units (yr$^{-1}$ magnetar$^{-1}$) and 90\% c.l. for the upper limits and, when needed, we also renormalised them to the same number of assumed Galactic magnetars (30).  
The black histogram shows the constraints derived using only the three confirmed MGFs in the Milky Way and LMC.  

\section{Conclusions}
\label{sec:conc}

We searched for MGFs in $\sim$35 Ms of INTEGRAL data on the Virgo cluster with negative results. With the reasonable assumption that the number of active magnetars in each galaxy scales with the rate of star formation, we derived a 90\% c.l. upper limit of about one giant flare with $E>3\times10^{45}$ erg every $\sim$500 yr per magnetar. We also put some constraints on the rate of MGFs with lower energies, thanks to INTEGRAL observations of a few nearby galaxies with high SFR. The analysis of this sample, with the detection of only one MGF in M82, implies a rate lower limit of  $R(>E)>4\times10^{-4}$ yr$^{-1}$ magnetar$^{-1}$, for energies $E<10^{45}$ erg.

These findings, in agreement with those obtained in similar analysis based on data from other satellites,   demonstrate the importance of enlarging the sample of MGFs through searches extending beyond the Local Group of galaxies. Although the detection of pulsating tails, which would unambiguously confirm the MGFs candidates, is difficult with the current instrumentation, the most recent discoveries   show  that quick and precise localisations with instruments providing good spectral/timing  capabilities coupled to rapid multiwavelength follow-ups are key elements to advance in this field.

\section*{Acknowledgements}
The  results reported in this article are based on data obtained
with INTEGRAL, an ESA mission with instruments and science data centres funded by ESA member states, and with the participation of the Russian Federation and the USA.
This work received financial support from INAF through the Magnetars Large Program Grant (PI S.Mereghetti).

 \section*{Data availability}
All the data used in this article are available in public archives.\\


\bibliographystyle{mnras}
\bibliography{main}

\bsp	
\label{lastpage}
\end{document}